\documentclass[final,5p,times,twocolumn]{elsarticle}
\usepackage{amssymb}
\usepackage{amsmath}
\usepackage{graphicx}
\usepackage{dcolumn}
\usepackage{bm}
\usepackage{hyperref}
\usepackage[mathlines]{lineno}
\usepackage{xcolor} 
\usepackage{amsthm}
\usepackage{natbib}
\usepackage[normalem]{ulem}
\usepackage{booktabs}
\usepackage{multirow}
\usepackage{threeparttable}

\usepackage{mathtools}

\begin{document}

\begin{frontmatter}

\title{Anharmonic Phonon Renormalization and Defect Tolerance of the Thermoelectric Power Factor in Monolayer SnSe}

\author[HCMUS,VNU]{Nguyen Tran Gia Bao}
\author[INOMAR,Health,VNU]{Thang Bach Phan}
\author[HCMUS,VNU]{Vu Thi Hanh Thu\corref{cor1}}
\ead{vththu@hcmus.edu.vn}
\cortext[cor1]{Corresponding author.}
\author[Tohoku]{Nguyen Tuan Hung\corref{cor2}}
\ead{nguyen.tuan.hung.e4@tohoku.ac.jp}
\cortext[cor2]{Corresponding author.}

\affiliation[HCMUS]{organization={Faculty of Physics and Physics Engineering, University of Science},
            city={Ho Chi Minh City},
            postcode={700000}, 
            country={Viet Nam}}

\affiliation[INOMAR]{
            organization={Advanced Materials Technology Institute Vietnam National University Ho Chi Minh City (formerly affiliated with Center for Innovative Materials and Architectures)},
            city={Ho Chi Minh City},
            postcode={700000}, 
            country={Viet Nam}}
            
\affiliation[Health]{
            organization={University of Health Sciences (UHS), Viet Nam National University Ho Chi Minh City},
            city={Ho Chi Minh City},
            postcode={700000}, 
            country={Viet Nam}}

\affiliation[VNU]{
            organization={Viet Nam National University Ho Chi Minh City},
            city={Ho Chi Minh City},
            postcode={700000}, 
            country={Viet Nam}}
\affiliation[Tohoku]{
            organization={Frontier Research Institute for Interdisciplinary Sciences, Tohoku University},
            city={Sendai},
            postcode={980-8578}, 
            country={Japan}}

\begin{abstract}
Monolayer tin selenide (SnSe) exhibits phase-dependent anharmonic lattice dynamics, yet their consequences for the thermoelectric power factor (PF) and point-defect tolerance remain unresolved. We combine density functional theory, the stochastic self-consistent harmonic approximation (SSCHA), and Boltzmann transport calculations including first-principles electron-phonon and electron-defect scattering to investigate monolayer $\alpha$-SnSe (\textit{Pnma}) and $\beta$-SnSe (\textit{Cmcm}). In dynamically stable $\alpha$-SnSe, SSCHA renormalizes the finite-temperature phonons without changing the qualitative $n$-type transport picture. Electron scattering is dominated by optical phonons, with TO-2 providing the largest resolved contribution at 100~K and LO-1 becoming dominant at 300~K. In $\beta$-SnSe, SSCHA removes the harmonic soft-mode instability of the \textit{Cmcm} phase at 800-1000~K, and thereby enables high-temperature transport calculations; LO/TO-2 is the principal electron-scattering channel. In the lower-density window near $10^{12}$~cm$^{-2}$, the $n$-type PF reaches 15-19~$\mu\mathrm{W}/(\mathrm{K}^{2}\cdot\mathrm{cm})$ at 800-900~K and exceeds the $p$-type PF primarily because of the higher electrical conductivity. Se vacancies ($V_{\mathrm{Se}}$) produce weaker electron-defect scattering than Sn vacancies ($V_{\mathrm{Sn}}$), and $p$-type transport is less defect tolerant than $n$-type transport in both phases. We define an operational critical defect concentration, $C_{\mathrm{crit}}$, at which the PF decreases by 15\% relative to the corresponding defect-free value. The lowest $C_{\mathrm{crit}}$ is $8.841\times10^{-5}$ (approximately 88~ppm) for $p$-type $\alpha$-SnSe with $V_{\mathrm{Sn}}$; for $n$-type $\beta$-SnSe with $V_{\mathrm{Se}}$, the 15\% threshold is not reached up to $5\times10^{-3}$ (5000~ppm). These results distinguish finite-temperature phonon renormalization in stable $\alpha$-SnSe from anharmonic stabilization in $\beta$-SnSe and provide defect-concentration limits for preserving the PF.

\end{abstract}





\end{frontmatter}



\section{Introduction}
Thermoelectric (TE) materials convert a temperature gradient directly into electrical energy and therefore offer a solid-state route for waste-heat recovery from industrial processes and electronic devices \cite{Hung2017-bj, Van-Thanh2023-ws, Hung2016-ll, Van-Thanh2025-ea, Fu2025-tq}. Their conversion efficiency is governed by the dimensionless figure of merit
$ZT = S^{2}\sigma T/(\kappa_{\mathrm{L}}+\kappa_{\mathrm{el}})$,
where $S$ is the Seebeck coefficient, $\sigma$ is the electrical conductivity,
and $\kappa_{\mathrm{L}}$ and $\kappa_{\mathrm{el}}$ represent lattice and electronic thermal conductivity, respectively, and $T$ is the operating temperature. Because $ZT$ increases with power factor $\text{PF} = S^{2}\sigma$ and decreases with total thermal conductivity $\kappa=\kappa_{\mathrm{L}}+\kappa_{\mathrm{el}}$, high-performance TE materials must combine a large PF with a low $\kappa$ \cite{Hung2019-hy}. 
For practical deployment, $ZT > 2$ is often used as a target for high-performance TE materials \cite{hung2021origin,li2010high}, whereas values approaching $ZT \sim 10$ are needed to compete with other energy technologies such as photovoltaics \cite{fu2025ai}.
$ZT$ and PF serve distinct roles at the device level: while $ZT$ determines the maximum conversion efficiency, PF determines the electrical power that can be produced for a given temperature gradient and device geometry.
\cite{Hung2016-ll, hung2021origin}. This distinction is particularly relevant in low-dimensional materials, where carrier-density control and electronic scattering can be engineered to optimize the PF independently of the lattice thermal conductivity.

Tin selenide (SnSe) is a representative anharmonic TE material with high $ZT$ over approximately 723-973~K \cite{Zhou2022-nr}. Bulk SnSe undergoes a structural transformation from low-temperature $\alpha$-SnSe (\textit{Pnma}) toward high-temperature $\beta$-SnSe (\textit{Cmcm}), beginning near 600~K and becoming complete near 800~K \cite{Zhou2022-nr, Hong2019-dr}. The high-temperature phase combines a reported PF of 10.1~$\mu\mathrm{W}/(\mathrm{K}^{2}\cdot\mathrm{cm})$ with a lattice thermal conductivity as low as 0.23~$\mathrm{W}/(\mathrm{K}\cdot\mathrm{m})$ \cite{Zhao2014-so}. The suppressed lattice heat transport originates from strong phonon anharmonicity associated with a bonding instability \cite{Zhou2022-nr, Hong2019-dr, Skelton2016-il, Chang2018-ac}. In bulk $\beta$-SnSe, this instability appears as a $\Gamma$-point ``ferroelectric-like'' soft transverse-optical mode, although the low-temperature bulk phase remains nonpolar \cite{Hong2019-dr}. A peak $ZT=2.6\pm0.3$ at 923~K has been reported for single-crystal SnSe \cite{Zhao2014-so}.

Monolayer SnSe has also been synthesized \cite{Li2022-bh, Wang2015-et, Chang2020-rh, Yue2024-oa}, motivating first-principles studies of its electronic and thermoelectric transport \cite{Ding2019-tf, Li2022-bh, Gupta2021-nm}. Although single-layer SnSe is often predicted to exhibit a larger lattice thermal conductivity and hence a lower intrinsic $ZT$ than bulk \cite{Ding2019-tf}, low-dimensional electronic-structure effects (e.g., quantum confinement, valley degeneracy tuning, and carrier-density control) can still yield competitive performance through PF optimization \cite{Hung2016-ll, hung2021origin}. Therefore, assessing the mechanisms that control PF in monolayer SnSe is essential for evaluating its practical potential in waste-heat harvesting. However, several unresolved challenges remain in the study of the TE properties of 2D SnSe. First, monolayer SnSe has been reported to be grown by molecular beam epitaxy (MBE) \cite{Yue2024-oa, Chang2020-rh}, where point defects and disorder can be introduced during synthesis and add extrinsic scattering channels \cite{Yue2024-oa}. These scattering channels are expected to reduce the carrier mobility, electrical conductivity, and the PF. Second, the effect of anharmonic lattice dynamics on the PF remains underexplored, despite its importance for accurate modeling of 2D TE materials \cite{Fei2016-ji}. Anharmonicity is not limited to imaginary phonon modes. Even in harmonically stable crystals, anharmonic renormalization can shift phonon frequencies and alter the mode-resolved scattering phase space, leading to non-negligible changes in e-ph scattering rates and transport coefficients \cite{Sun2025-fm, Ribeiro_undated-sq, Zhou2018-gh, Ranalli2024-iu}.

We examine these effects in the low-temperature $\alpha$-SnSe phase and the high-temperature $\beta$-SnSe phase. For $\alpha$-SnSe, whose harmonic phonon spectrum is dynamically stable, SSCHA provides finite-temperature-renormalized phonons at 100 and 300~K. For $\beta$-SnSe, whose harmonic spectrum contains imaginary-frequency modes, SSCHA stabilizes the lattice at 800, 900, and 1000~K before electron-phonon transport is evaluated. We resolve the scattering rates by phonon mode to identify the dominant branches in each phase. We then combine electron-phonon and electron-defect transport contributions and define an operational critical concentration, $C_{\mathrm{crit}}$, below which the PF remains within 15\% of its phonon-limited value. The resulting bounds quantify the vacancy concentrations compatible with preserving the PF of monolayer SnSe.

\section{Results and Discussion}

\subsection{Crystal, Electronic, and Vibrational Structures}

\subsubsection{Structural phases: $\alpha$-SnSe (\textit{Pnma}) and $\beta$-SnSe (\textit{Cmcm})}
\begin{figure*}[!t]
    \centering
    \includegraphics[width=0.9\linewidth]{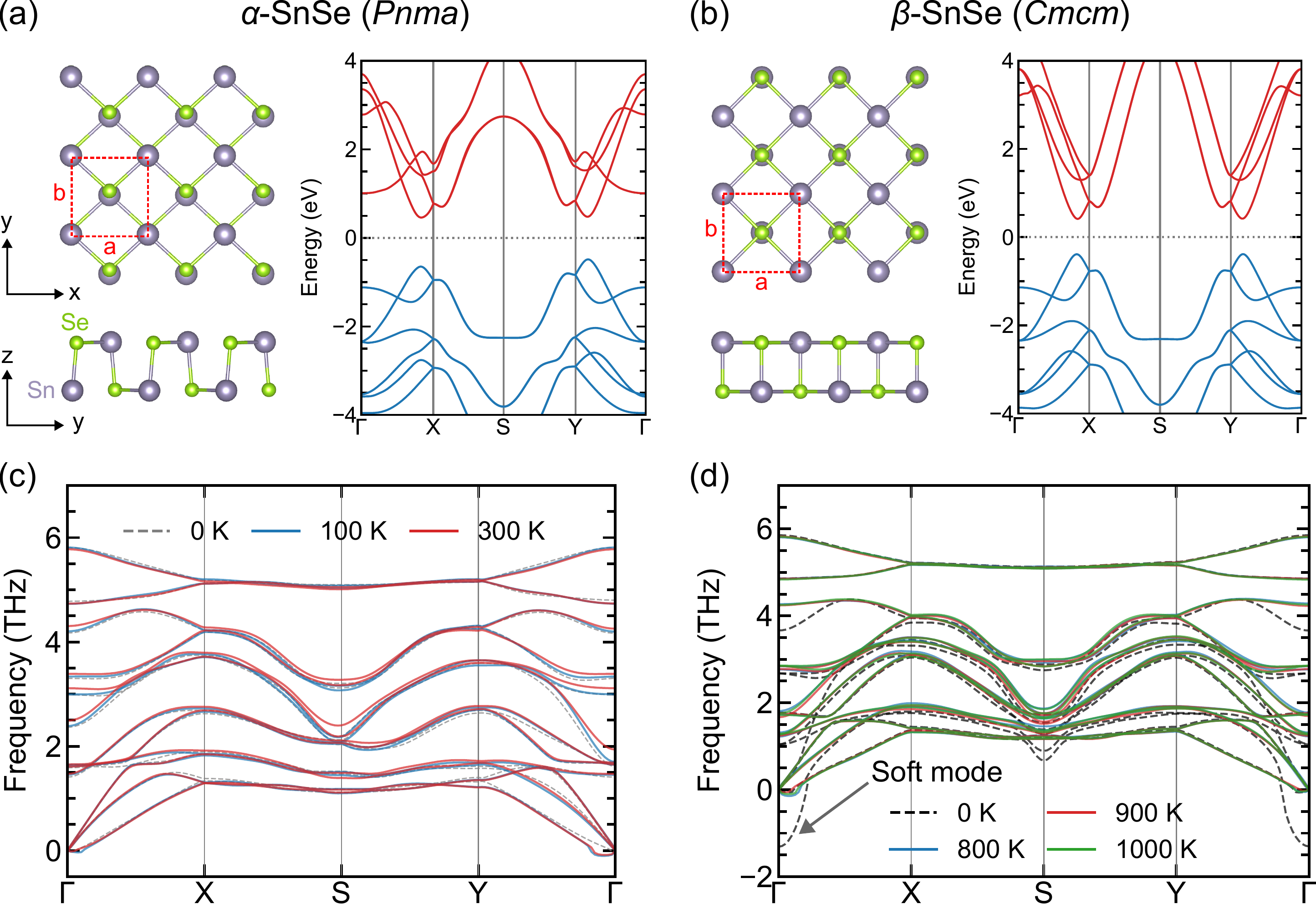}
    \caption{Relaxed structures, Wannier-interpolated electronic bands, and phonon dispersions of monolayer SnSe. Panels (a) and (b) show top and side views of the relaxed $\alpha$- and $\beta$-SnSe structures together with the corresponding band structures along $\Gamma$-X-S-Y-$\Gamma$. Sn atoms are light purple and Se atoms are light green. Panels (c) and (d) compare the harmonic DFPT phonons with the SSCHA-renormalized phonons for $\alpha$-SnSe at 100 and 300~K and for $\beta$-SnSe at 800, 900, and 1000~K, respectively.}

    \label{fig:crystal-and-band-phonon}
\end{figure*}

At room temperature, bulk SnSe adopts a layered orthorhombic structure with space group \textit{Pnma}. This structure is a low-symmetry distortion of the higher-symmetry \textit{Cmcm} arrangement, which is related to a rocksalt-type lattice \cite{Wang2015-et}. Fig.~\ref{fig:crystal-and-band-phonon} shows the corresponding monolayer structures. The \textit{Pnma}-like monolayer breaks inversion symmetry and supports an in-plane ferroelectric polarization, whereas the \textit{Cmcm}-like monolayer retains inversion symmetry and is nonpolar \cite{Fei2016-ji, Chang2020-rh}.

The stable phase depends on temperature. At ambient pressure, bulk SnSe transforms from the low-symmetry \textit{Pnma} phase to the higher-symmetry \textit{Cmcm} phase at a critical temperature of approximately 800~K \cite{Aseginolaza2019-jp,Zhao2014-so}. For a monolayer, first-principles calculations predict a lower polar-to-nonpolar transition temperature, $T_\mathrm{c}=325$~K \cite{Fei2016-ji}. Therefore, near room temperature the monolayer can fall on either side of $T_\mathrm{c}$, which leads to distinct lattice dynamics and allowed carrier-scattering channels in the polar \textit{Pnma} structure and the nonpolar \textit{Cmcm} structure. To isolate these effects, we analyze the electronic structure and transport response of the two phases separately.

After geometry optimization, the $\alpha$-SnSe monolayer in the \textit{Pnma} phase has an orthorhombic unit cell with in-plane lattice constants $a=4.29$~\AA{} and $b=4.40$~\AA{}, where $a$ and $b$ are defined as shown in Fig.~\ref{fig:crystal-and-band-phonon}. These values agree with previous theoretical and experimental results~\cite{Fei2016-ji, Fei2015-sn}.
The calculated monolayer lattice constants are close to the corresponding bulk values, indicating similar in-plane bonding geometries in the two forms \cite{Fei2016-ji}. The $\alpha$-SnSe monolayer has a puckering height of $h=2.76$~\AA{} and zigzag Sn-Se chains along the $y$ direction (Fig.~\ref{fig:crystal-and-band-phonon}). The $\beta$-SnSe monolayer is also orthorhombic but has an approximately square in-plane lattice, $a=b=4.31$~\AA{}, and a puckering height of $h=2.73$~\AA{}. This phase is more symmetric, with a hinge angle closer to the high-symmetry limit, resulting in a weaker out-of-plane puckering.

For the 2D Fr\"ohlich correction to polar e-ph scattering and the conversion of sheet conductance to volume-normalized conductivity, we use the effective monolayer thickness
\begin{equation*}
    t = d_{\mathrm{outer}} + 2r_{\mathrm{vdW}}^{\max},
\end{equation*}
where $d_{\mathrm{outer}}$ is the vertical distance between the two outermost atoms, and $r_{\mathrm{vdW}}^{\max}$ is the larger van der Waals radius among the surface atoms \cite{Bao2025-gv}. Because both surfaces of monolayer SnSe contain Sn and Se atoms, we take $r_{\mathrm{vdW}}^{\max}=2.42$~\AA{} \cite{alvarez2013cartography}. With $d_{\mathrm{outer}}=2.75$~\AA{}, we obtain $t=7.59$~\AA{}, within the experimental thickness range of $6.8\pm1.4$~\AA{} \cite{Jiang_2017, xue2022layer}.

\subsubsection{Band structure and valley alignment}
Figs.~\ref{fig:crystal-and-band-phonon}(a) and (b) present the Wannier-interpolated band structures of $\alpha$- and $\beta$-SnSe, respectively. The orthorhombic Brillouin-zone path and a comparison between the DFT and Wannier-interpolated bands are provided in Fig.~S1 (Supporting Information). The interpolated bands agree with the DFT bands within the energy window used for transport. Both phases are indirect-gap semiconductors, with calculated gaps of $E_g=0.94$~eV for $\alpha$-SnSe and $E_g=0.80$~eV for $\beta$-SnSe, within the range of previous reports \cite{Fei2015-sn, Zhao2014-so}. In both phases, the VBM lies along Y-$\Gamma$, whereas the CBM lies along $\Gamma$-X.

Fig.~S1 (Supporting Information) shows secondary valleys near both band edges. The secondary VBM lies along $\Gamma$-X, and the secondary CBM lies along Y-$\Gamma$. In $\alpha$-SnSe, the secondary VBM is 163.24~meV below the VBM and the secondary CBM is 19.13~meV above the CBM. The corresponding offsets in $\beta$-SnSe are 2.21 and 6.07~meV. These few-meV offsets indicate near-degenerate band-edge valleys in the higher-symmetry $\beta$-SnSe phase. Table~S1 (Supporting Information) lists the four band-edge energies.

\subsubsection{Harmonic and finite-temperature lattice dynamics}

The lattice dynamics of the two structural phases differ qualitatively, as shown in Fig.~\ref{fig:crystal-and-band-phonon}(c) for $\alpha$-SnSe and Fig.~\ref{fig:crystal-and-band-phonon}(d) for $\beta$-SnSe. For $\alpha$-SnSe, the harmonic DFPT phonon dispersion in Fig.~\ref{fig:crystal-and-band-phonon}(c) contains no imaginary frequencies along the high-symmetry path, confirming that the relaxed monolayer is dynamically stable within the harmonic approximation. Finite-temperature SSCHA calculations renormalize both the acoustic and optical branches. Relative to the harmonic result, most acoustic branches soften, whereas the optical branches harden. The magnitude of these frequency shifts varies with temperature, but no dynamical instability emerges. Thus, for $\alpha$-SnSe, SSCHA accounts for finite-temperature phonon-frequency renormalization rather than stabilizing the crystal structure. The corresponding phonon dispersions, mode assignment, and representative phonon eigenvectors are provided in Figs.~S2-S4 and Table~S2 (Supporting Information).

By contrast, the harmonic DFPT phonon dispersion of $\beta$-SnSe in Fig.~\ref{fig:crystal-and-band-phonon}(d) contains imaginary-frequency branches, indicating that the high-symmetry \textit{Cmcm} structure is dynamically unstable within the zero-temperature harmonic approximation. This instability is dominated by a soft transverse optical (TO) mode associated with the lattice dynamics of the high-temperature phase. Finite-temperature anharmonic renormalization within SSCHA stabilizes this mode and removes the imaginary frequencies at 800, 900, and 1000~K, as shown in Fig.~\ref{fig:crystal-and-band-phonon}(d). The harmonic and SSCHA dispersions nevertheless remain similar for the higher-frequency branches above approximately 3~THz, indicating that the largest anharmonic correction is concentrated in the low-frequency soft-mode region. The SSCHA-renormalized phonons are therefore required for the subsequent electron-phonon transport calculations of $\beta$-SnSe, whereas in $\alpha$-SnSe they renormalize an already dynamically stable phonon spectrum. The SSCHA-renormalized phonon dispersion, mode assignment, and representative phonon eigenvectors of $\beta$-SnSe are provided in Figs.~S12-S13 and Table~S4 (Supporting Information).

\subsection{Effects of Anharmonic Phonon Renormalization on Transport in $\alpha$-SnSe}

\begin{figure*}[!t]
    \centering
    \includegraphics[width=0.8\linewidth]{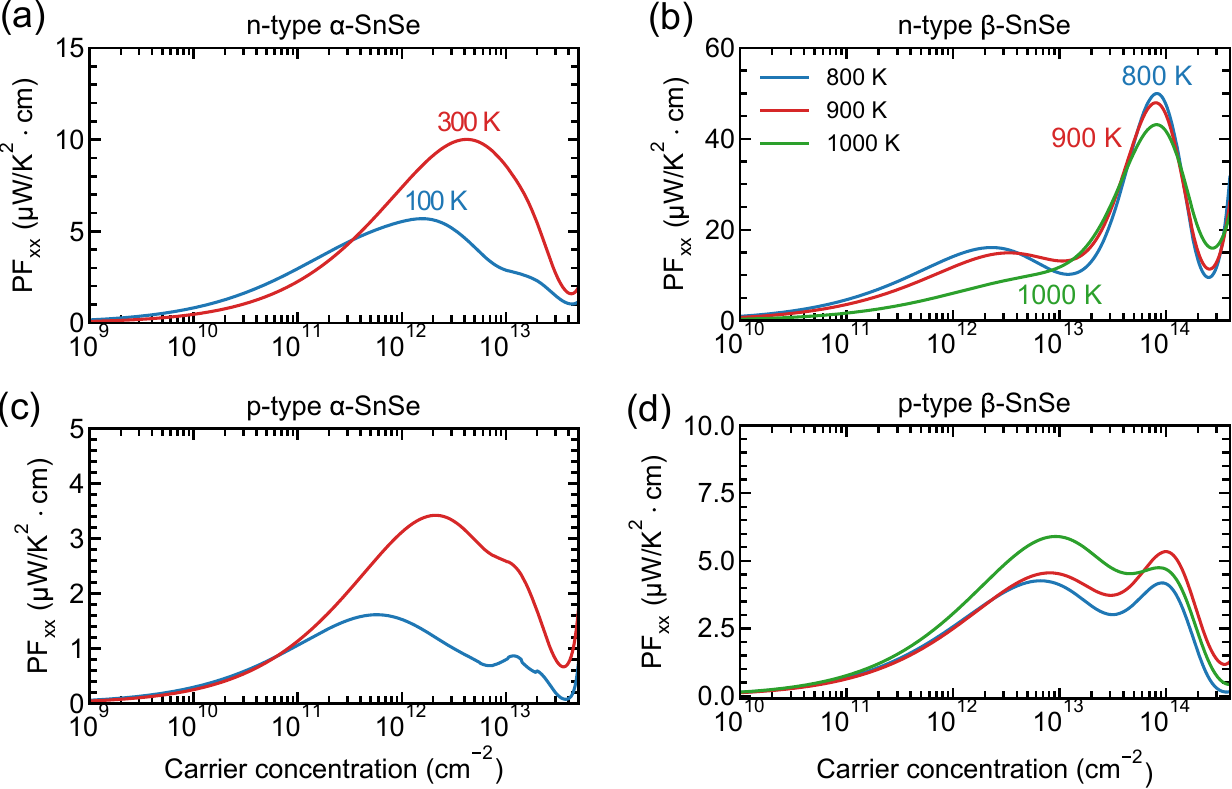}
    \caption{Power factor along the $xx$ direction ($\mathrm{PF}_{xx}$) as a function of carrier concentration, calculated using SSCHA-renormalized phonons. Panels (a) and (c) show the $n$- and $p$-type results for monolayer $\alpha$-SnSe at 100 and 300 K, respectively, while panels (b) and (d) show the corresponding results for $\beta$-SnSe at 800, 900, and 1000 K.}
    \label{fig:PF-alpha-beta}
\end{figure*}

Because the harmonic phonon spectrum of $\alpha$-SnSe is already dynamically stable, SSCHA is used here as a finite-temperature correction rather than as a prerequisite for stabilizing the structure. The harmonic and SSCHA calculations use the same electronic band structure and therefore yield similar Seebeck coefficients over most of the carrier-concentration range, as shown in Fig.~S8-S11 (Supporting Information). Differences in the power factor arise mainly from phonon-induced changes in the carrier lifetime and electrical conductivity.

For $n$-type $\alpha$-SnSe, the first local PF maximum remains on the same overall scale after phonon renormalization. Along $xx$, it changes from 6.827 to 5.688~$\mu\mathrm{W}/(\mathrm{K}^{2}\cdot\mathrm{cm})$ at 100~K and from 8.963 to 10.014~$\mu\mathrm{W}/(\mathrm{K}^{2}\cdot\mathrm{cm})$ at 300~K. Along $yy$, the corresponding changes are from 7.382 to 6.954 and from 8.958 to 9.112~$\mu\mathrm{W}/(\mathrm{K}^{2}\cdot\mathrm{cm})$. Fig.~\ref{fig:PF-alpha-beta}(a) further shows that the SSCHA-based $\mathrm{PF}_{xx}$ remains higher at 300~K than at 100~K and that its maximum shifts to a larger carrier concentration. Thus, phonon renormalization changes the numerical PF values without changing the qualitative magnitude or temperature trend of the $n$-type response. Table~S3 (Supporting Information) lists the PF-maximizing carrier concentrations and the corresponding $S$, $\sigma$, and PF values for both carrier types and transport directions.

This behavior is consistent with the conduction-band electron-phonon scattering rates in Fig.~S5 (Supporting Information). At 100 and 300~K, the acoustic, optical, and total rates obtained with the harmonic and SSCHA phonons remain similar over the transport-relevant energy range. In both cases, electron-phonon scattering is dominated by optical phonons, whereas the acoustic contribution is lower. SSCHA therefore primarily redistributes the scattering contributions among individual phonon modes without markedly changing the total scattering rate.

Figure~\ref{fig:scattering-rates}(a) resolves the SSCHA total scattering rate into contributions from the LO-1, LO-2, LO-3, TO-1, TO-2, TO-3, and ZO-1 modes. These branches provide the largest optical-mode contributions; the remaining branches are shown in Fig.~S6 (Supporting Information). At 100~K, TO-2 contributes the largest rate over most of the transport-relevant electron-energy range. At 300~K, the LO-1 contribution increases and becomes the dominant resolved channel. The temperature dependence of the total scattering rate therefore reflects unequal increases among the optical branches, with the larger increase of LO-1 scattering causing it to become the dominant resolved channel at 300~K.

The $p$-type PF is more sensitive to phonon renormalization. Along $xx$, the maximum changes from 4.439 to 1.615~$\mu\mathrm{W}/(\mathrm{K}^{2}\cdot\mathrm{cm})$ at 100~K and from 5.757 to 3.422~$\mu\mathrm{W}/(\mathrm{K}^{2}\cdot\mathrm{cm})$ at 300~K. Along $yy$, it changes from 4.822 to 1.803 and from 2.651 to 2.814~$\mu\mathrm{W}/(\mathrm{K}^{2}\cdot\mathrm{cm})$, respectively. Fig.~\ref{fig:PF-alpha-beta}(c) shows that the SSCHA-based $p$-type $\mathrm{PF}_{xx}$ maximum at 300~K is more than twice its value at 100~K. Figs.~S10-S11 and Table~S3 (Supporting Information) provide the full carrier-concentration dependence and the associated $S$ and $\sigma$ values.

Overall, SSCHA acts as a finite-temperature correction for $\alpha$-SnSe rather than as a mechanism required to stabilize the structure. For $n$-type transport, phonon renormalization changes the relative contributions of individual optical modes, including a change in the dominant mode from TO-2 at 100~K to LO-1 at 300~K. However, the total electron-phonon scattering rate remains similar to the harmonic result, so the PF values are modified without changing their overall magnitude or temperature trend. In contrast, the $p$-type PF shows a stronger quantitative sensitivity to phonon renormalization.

\subsection{Soft-Mode Stabilization and Transport in $\beta$-SnSe}

\begin{figure*}[!t]
    \centering
    \includegraphics[width=0.8\linewidth]{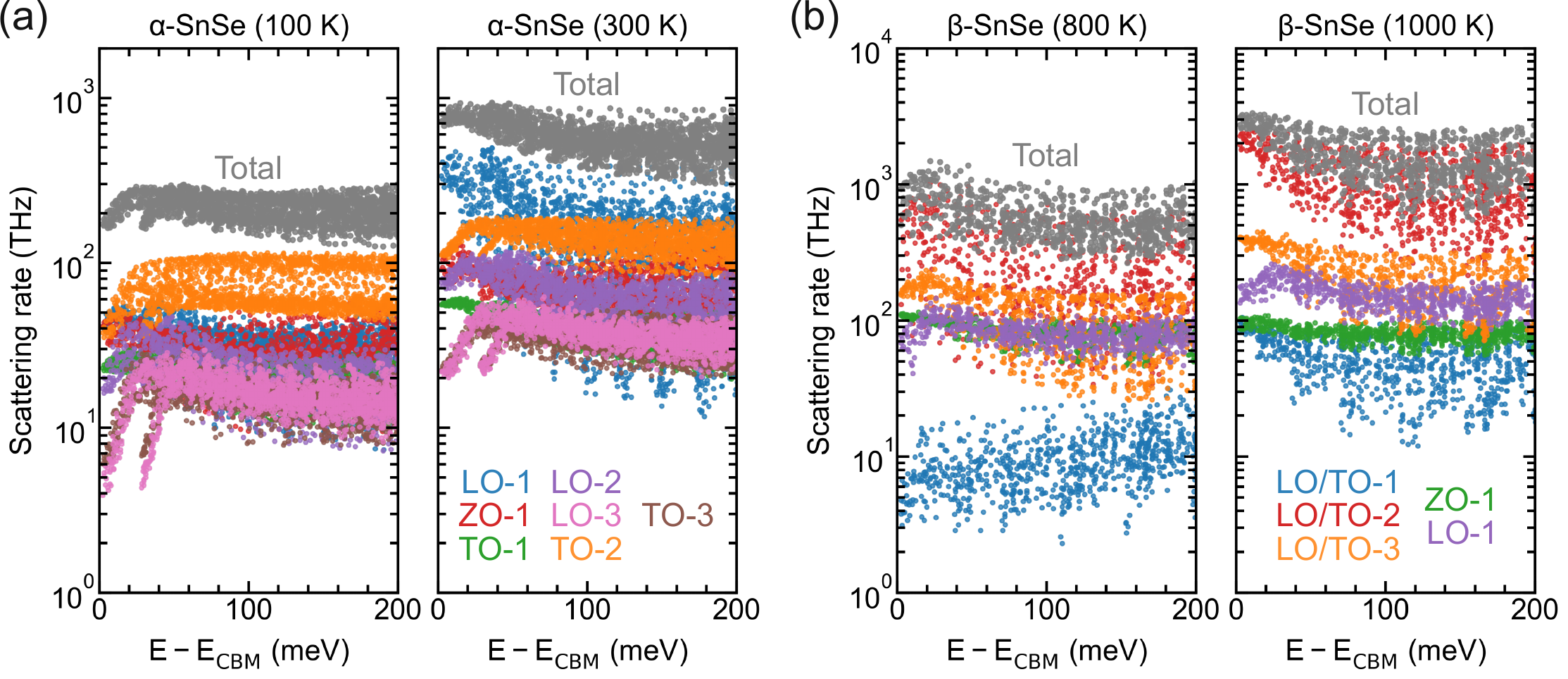}
    \caption{Mode-resolved electron-phonon scattering rates as a function of conduction-band energy, calculated using SSCHA-renormalized phonons for (a) $\alpha$-SnSe at 100 and 300~K and (b) $\beta$-SnSe at 800 and 1000~K. Panel (a) shows LO-1, LO-2, LO-3, ZO-1, TO-1, TO-2, and TO-3; panel (b) shows LO/TO-1, LO/TO-2, LO/TO-3, ZO-1, and LO-1. Gray points are the total scattering rates summed over all 12 phonon modes.}

    \label{fig:scattering-rates}
\end{figure*}

We next analyze conduction-band electron-phonon scattering in $\beta$-SnSe using SSCHA-renormalized phonons. Figure~\ref{fig:scattering-rates}(b) shows that a small subset of optical branches dominates the scattering rate. LO/TO-2 provides the largest contribution at both 800 and 1000~K, followed by LO/TO-3 and LO-1, whereas ZO-1 and LO/TO-1 contribute less. This hierarchy remains unchanged between the two temperatures, although the LO/TO-2 rate increases markedly and reaches values of approximately $10^{3}$~THz at 1000~K within the investigated energy range. The temperature-induced increase in the total scattering rate therefore arises primarily from stronger LO/TO-2 scattering rather than from a change in the dominant branch. The remaining branches are generally weaker, as shown in Fig.~S7 (Supporting Information). SSCHA also stabilizes the soft TO-1 mode, whose scattering rate increases with temperature but remains below those of LO/TO-2, LO/TO-3, and LO-1. These results identify LO/TO-2 as the dominant resolved optical-phonon scattering channel in high-temperature electron transport in monolayer $\beta$-SnSe.

Having identified the dominant scattering channels, we next assess their consequences for the thermoelectric transport of monolayer $\beta$-SnSe. The full carrier-concentration dependence of the Seebeck coefficient, electrical conductivity, and power factor along both $xx$ and $yy$ directions is provided in Figs.~S14-S17 (Supporting Information). As shown in Figs.~\ref{fig:PF-alpha-beta}(b) and (d), $n$-type doping yields a higher PF than $p$-type doping in the lower carrier-concentration range near $10^{12}$~cm$^{-2}$. At 800~K, the first local $n$-type PF maxima are 16.090 and 19.083~$\mu\mathrm{W}/(\mathrm{K}^{2}\cdot\mathrm{cm})$ along the $xx$ and $yy$ directions, respectively. These maxima occur at electron concentrations of $2.34\times10^{12}$ and $2.08\times10^{12}$~cm$^{-2}$. At 900~K, the corresponding maxima decrease to 14.920 and 15.107~$\mu\mathrm{W}/(\mathrm{K}^{2}\cdot\mathrm{cm})$, respectively. They occur at $3.37\times10^{12}$ and $2.82\times10^{12}$~cm$^{-2}$, respectively. At 1000~K, no distinct local maximum is found below $5\times10^{12}$~cm$^{-2}$. At the upper boundary of this low-density window, the PF values are 9.800 and 2.444~$\mu\mathrm{W}/(\mathrm{K}^{2}\cdot\mathrm{cm})$ along the $xx$ and $yy$ directions, respectively. Because no local peak occurs within this window, we report these values as reference points rather than as true local maxima.

The $p$-type PF remains lower, with local maxima ranging from 3.963 to 8.206~$\mu\mathrm{W}/(\mathrm{K}^{2}\cdot\mathrm{cm})$ at hole concentrations of approximately $(6.4$-$9.1)\times10^{12}$~cm$^{-2}$. At the $n$- and $p$-type local maxima and the 1000~K $n$-type reference points, the Seebeck coefficients of the two doping types are comparable, generally ranging from 0.168 to 0.187~mV/K. The higher $n$-type PF therefore originates mainly from the larger electrical conductivity. The $n$-type conductivity reaches approximately $(2.8$-$5.4)\times10^{4}$~S/m, compared with $(1.4$-$2.7)\times10^{4}$~S/m for $p$-type doping. Thus, the $n$-type advantage in this carrier-concentration range reflects the higher electrical conductivity rather than a larger Seebeck coefficient.

At higher electron concentrations, the global PF maxima reach approximately 43-50~$\mu\mathrm{W}/(\mathrm{K}^{2}\cdot\mathrm{cm})$. Along the $xx$ direction, the global maximum decreases from 49.948~$\mu\mathrm{W}/(\mathrm{K}^{2}\cdot\mathrm{cm})$ at 800~K to 47.911 and 43.098~$\mu\mathrm{W}/(\mathrm{K}^{2}\cdot\mathrm{cm})$ at 900 and 1000~K, respectively. Along the $yy$ direction, the corresponding values are 49.049, 47.122, and 46.968~$\mu\mathrm{W}/(\mathrm{K}^{2}\cdot\mathrm{cm})$. These maxima occur at electron concentrations of approximately $(8.0$-$8.9)\times10^{13}$~cm$^{-2}$. Their large PF values arise from electrical conductivities of $(3.1$-$6.2)\times10^{5}$~S/m, which compensate for the smaller Seebeck coefficients of 0.089-0.118~mV/K. With increasing temperature, stronger electron-phonon scattering shortens the carrier relaxation times and reduces the electrical conductivity. This reduction primarily accounts for the decrease in the global $\mathrm{PF}_{xx}$ maximum. By contrast, the $\mathrm{PF}_{yy}$ maximum decreases less and slightly exceeds $\mathrm{PF}_{xx}$ at 1000~K.

Carrier densities approaching $10^{14}$~cm$^{-2}$ can, in principle, be induced in two-dimensional materials through high-capacitance electrostatic \cite{song2022high} or ionic gating \cite{weintrub2022generating}. However, sustaining such carrier densities using conventional ionic-liquid or ion-gel gating would be experimentally difficult at 800-1000~K. Reaching a comparable carrier density through chemical doping would likely make the material degenerately doped. At such doping levels, band filling and dopant-induced changes to the electronic structure may reduce the validity of the rigid-band approximation. We therefore interpret the high-density maxima as phonon-limited upper bounds of the present transport model. These maxima should be distinguished from the lower-density PF features near $10^{12}$~cm$^{-2}$. All numerical values are summarized in Tables~S5 and S6 (Supporting Information).

\subsection{Point-Defect Scattering and Defect Tolerance}

\begin{figure*}
    \centering
    \includegraphics[width=1\linewidth]{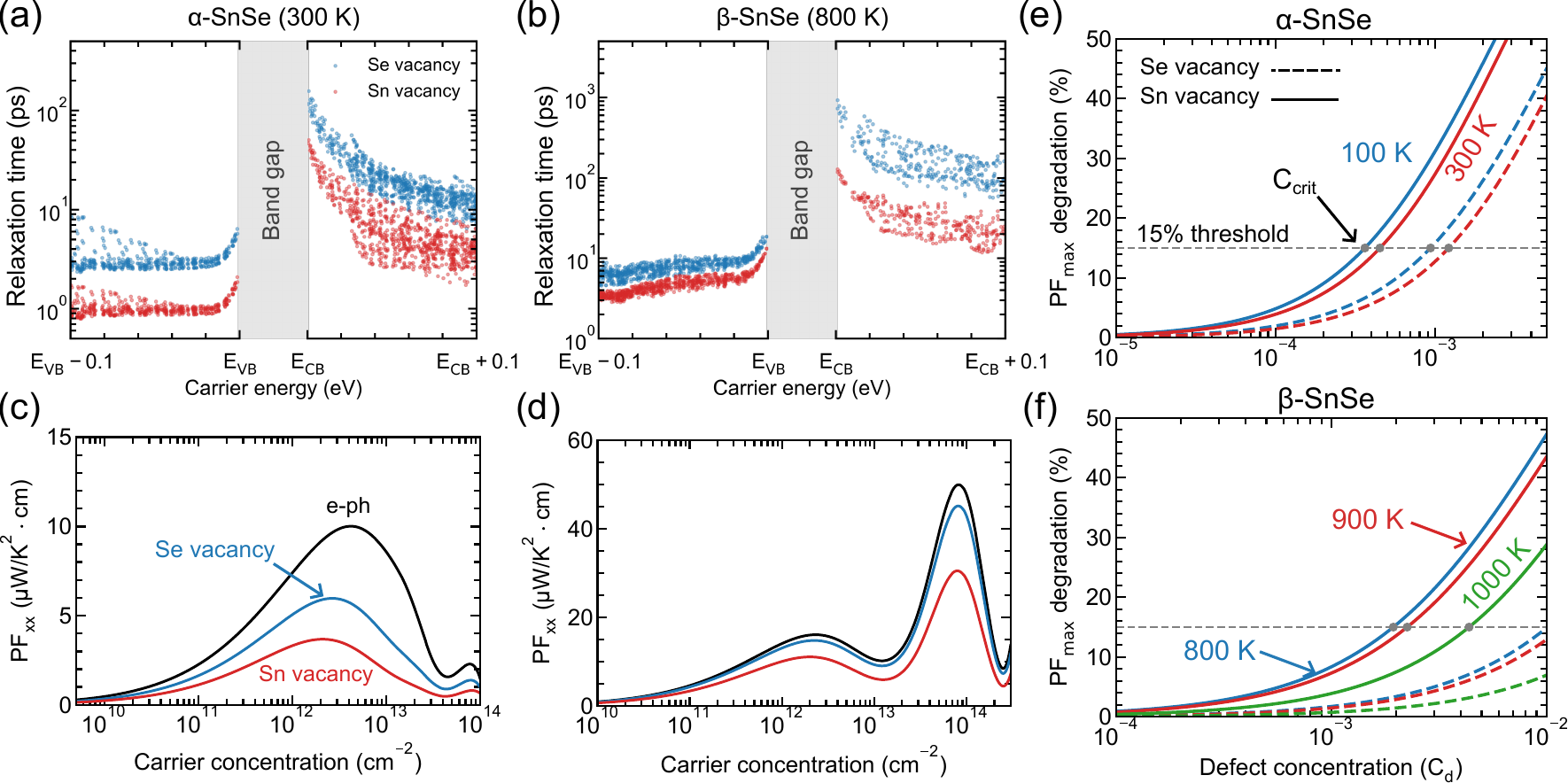}
    \caption{Electron-defect relaxation time as a function of carrier energy for (a) $\alpha$-SnSe at 300~K and (b) $\beta$-SnSe at 800~K, evaluated at 1~ppm. Power factor along $xx$ ($\mathrm{PF}_{xx}$) as a function of two-dimensional electron concentration for (c) $\alpha$-SnSe and (d) $\beta$-SnSe at $C_d=5\times10^{-3}$ (5000~ppm). PF degradation as a function of $C_d$ for (e) $\alpha$-SnSe and (f) $\beta$-SnSe. The horizontal dashed line marks the operational 15\% threshold, and $C_{\mathrm{crit}}$ is the corresponding critical concentration.}
    \label{fig:fig6}
\end{figure*}

\begin{table*}[htb]
\centering
\caption{Critical defect concentration $C_{\mathrm{crit}}$ defined by the operational 15\% PF-degradation threshold along the $xx$ direction for $n$- and $p$-type monolayer SnSe. For $n$-type $\beta$-SnSe at 1000~K, degradation is evaluated at a fixed carrier concentration of $5\times10^{12}$~cm$^{-2}$.}

\label{tab:ccrit}
\begin{threeparttable}
\begin{tabular}{lccccc}
\toprule
\multirow{2}{*}{Phase}
& \multirow{2}{*}{$T$ (K)}
& \multicolumn{2}{c}{$n$-type ; $C_{\mathrm{crit}}$}
& \multicolumn{2}{c}{$p$-type ; $C_{\mathrm{crit}}$} \\
\cmidrule(lr){3-4}\cmidrule(lr){5-6}
&
& $V_{\mathrm{Sn}}$
& $V_{\mathrm{Se}}$
& $V_{\mathrm{Sn}}$
& $V_{\mathrm{Se}}$ \\
\midrule
\multirow{2}{*}{$\alpha$-SnSe}
& 100  & $3.631\times10^{-4}$ & $9.371\times10^{-4}$
& $8.841\times10^{-5}$ & $2.626\times10^{-4}$ \\
& 300  & $4.511\times10^{-4}$ & $1.217\times10^{-3}$
& $1.425\times10^{-4}$ & $4.276\times10^{-4}$ \\
\midrule
\multirow{3}{*}{$\beta$-SnSe}
& 800  & $1.941\times10^{-3}$ & $>5\times10^{-3}$
& $2.016\times10^{-3}$ & $3.147\times10^{-3}$ \\
& 900  & $2.251\times10^{-3}$ & $>5\times10^{-3}$
& $2.130\times10^{-3}$ & $3.270\times10^{-3}$ \\
& 1000 & $4.366\times10^{-3}$ & $>5\times10^{-3}$
& $1.738\times10^{-3}$ & $2.625\times10^{-3}$ \\
\bottomrule
\end{tabular}
\end{threeparttable}
\end{table*}

\subsubsection{Defect type and temperature dependence}

Beyond phonon scattering, intrinsic vacancies perturb the crystal potential and provide an additional elastic channel for carrier relaxation. We focus on Sn ($V_{\mathrm{Sn}}$) and Se ($V_{\mathrm{Se}}$) vacancies because both have been identified as important intrinsic point defects in SnSe. $V_{\mathrm{Sn}}$ is closely associated with intrinsic $p$-type conduction, whereas the role of $V_{\mathrm{Se}}$ depends on temperature and the Fermi-level position \cite{sraitrova2019vacancies,wei2018achieving}. In particular, $V_{\mathrm{Sn}}$ have been identified as the primary origin of intrinsic $p$-type conduction in SnSe \cite{nguyen2022unidentified}. We evaluate how the two vacancy species alter the relaxation time and PF when combined with the phonon-limited transport described above.

Figs.~\ref{fig:fig6}(a) and (b) show the electron-defect (e-d) relaxation times at a defect concentration of $C_d=10^{-6}$ (1~ppm) for $\alpha$-SnSe and $\beta$-SnSe, respectively. $V_{\mathrm{Se}}$ yields longer relaxation times than $V_{\mathrm{Sn}}$ for both electron and hole transport in both structures, indicating weaker e-d scattering. Fig.~S18 (Supporting Information) shows that this ordering persists over the temperature range.

We then evaluate the effect of e-d scattering on the power factor using the expression derived from Eq.~\eqref{eq:PF_Cd}. In this expression, the Seebeck coefficient is assumed to remain nearly unchanged, as verified in Fig.~S19 (Supporting Information). Figs.~\ref{fig:fig6}(c) and (d) show the resulting $\mathrm{PF}_{xx}$ values at a defect concentration of $C_d=5\times10^{-3}$, corresponding to 5000 ppm. For $\alpha$-SnSe at 300~K, Fig.~\ref{fig:fig6}(c) shows that the peak $\mathrm{PF}_{xx}$ decreases from the electron-phonon-limited value of $10.014~\mu\mathrm{W}/(\mathrm{K}^{2}\cdot\mathrm{cm})$ to $5.969~\mu\mathrm{W}/(\mathrm{K}^{2}\cdot\mathrm{cm})$ for $V_{\mathrm{Se}}$ and to $3.681~\mu\mathrm{W}/(\mathrm{K}^{2}\cdot\mathrm{cm})$ for $V_{\mathrm{Sn}}$. This ordering follows the longer e-d relaxation time obtained for $V_{\mathrm{Se}}$. The same ordering is obtained for $\beta$-SnSe at 800~K, where the peak $\mathrm{PF}_{xx}$ decreases from $16.090~\mu\mathrm{W}/(\mathrm{K}^{2}\cdot\mathrm{cm})$ in the electron-phonon-limited case to $14.798~\mu\mathrm{W}/(\mathrm{K}^{2}\cdot\mathrm{cm})$ for $V_{\mathrm{Se}}$ and $11.091~\mu\mathrm{W}/(\mathrm{K}^{2}\cdot\mathrm{cm})$ for $V_{\mathrm{Sn}}$. Because defect scattering reduces the electrical conductivity while the Seebeck coefficient remains unchanged, the maximum power factor shifts toward lower carrier concentrations.

\subsubsection{Critical defect concentration for PF degradation}

Having established the dependence on defect type at fixed $C_d=5\times10^{-3}$ (5000~ppm), we now quantify how $\mathrm{PF}_{\max}$ degrades across the full concentration range studied.
We restrict the analysis to $C_d \leq 5\times10^{-3}$, within the
dilute-defect limit of the \citeauthor{Lu2019-ch} formalism, where defects
scatter independently and the pristine host band structure is
unperturbed~\cite{Lu2019-ch, Lu2020-cv}.
We define the PF degradation as
\begin{equation*}
\mathrm{PF}_{\mathrm{deg}}(C_d)
=
\frac{
\mathrm{PF}_{\max}^{\mathrm{e\text{-}ph}}
-
\mathrm{PF}_{\max}^{\mathrm{e\text{-}ph+e\text{-}d}}(C_d)
}{
\mathrm{PF}_{\max}^{\mathrm{e\text{-}ph}}
}
\times 100\% .
\label{eq:pf_degradation}
\end{equation*}

We adopt 15\% as an operational degradation threshold for defect tolerance; the intersection of each $\mathrm{PF}_{\mathrm{deg}}$ curve with this threshold defines the critical defect concentration $C_{\mathrm{crit}}$, with a larger value indicating higher defect tolerance.
The results discussed below correspond to transport along the $xx$ direction. The $n$-type results are shown in Figs.~\ref{fig:fig6}(e) and~(f), and all $C_{\mathrm{crit}}$ values for both carrier types are collected in Table~\ref{tab:ccrit}. For $\beta$-SnSe at 1000~K, where no distinct low-density PF maximum is observed, the PF degradation is instead evaluated at a fixed carrier concentration of $5\times10^{12}$~cm$^{-2}$.
 
For $n$-type $\alpha$-SnSe, $V_{\mathrm{Se}}$ reaches the threshold at a
concentration consistently higher than $V_{\mathrm{Sn}}$, reflecting its longer
e-d relaxation times. Stronger e-ph scattering at 300~K reduces the fractional weight of e-d
scattering in the total resistivity, raising $C_{\mathrm{crit}}$ for both
vacancy types relative to 100~K. Specifically, $C_{\mathrm{crit}}$ increases from $3.631\times10^{-4}$ to $4.511\times10^{-4}$ for $V_{\mathrm{Sn}}$ and from $9.371\times10^{-4}$ to $1.217\times10^{-3}$ for $V_{\mathrm{Se}}$.

For $n$-type $\beta$-SnSe with $V_{\mathrm{Sn}}$, $C_{\mathrm{crit}}$ is $1.941\times10^{-3}$ at 800~K and $2.251\times10^{-3}$ at 900~K. At 1000~K, the fixed-concentration definition gives $4.366\times10^{-3}$. Because the 1000~K value is evaluated at $5\times10^{12}$~cm$^{-2}$ rather than at a local PF maximum, it should not be interpreted as a strictly equivalent continuation of the peak-based temperature trend. Fig.~\ref{fig:fig6}(f) further shows that $V_{\mathrm{Se}}$ does not reach the 15\% threshold at any of the three temperatures. Its electron-defect scattering is therefore insufficient to produce 15\% degradation within the investigated dilute-defect range.

For $p$-type SnSe, $C_{\mathrm{crit}}$ is generally lower than the corresponding $n$-type values, indicating a greater sensitivity of hole transport to point-defect scattering (Table~\ref{tab:ccrit} and Fig.~S20, Supporting Information). In $\alpha$-SnSe, the $p$-type values are approximately 2.8-4.1 times lower than the corresponding $n$-type values, consistent with a relatively stronger contribution of e-d scattering to the degradation of hole transport near the valence-band edge. As illustrated in Fig.~S20 (Supporting Information), $C_{\mathrm{crit}}$ increases from 100 to 300~K for both vacancy types, consistent with the $n$-type $\alpha$-SnSe trend. The rise of $C_{\mathrm{crit}}$ is expected from the increasing contribution of e-ph resistivity at higher temperature, which reduces the fractional contribution of e-d scattering to the total resistivity.

In $\beta$-SnSe, both $V_{\mathrm{Sn}}$ and $V_{\mathrm{Se}}$ reach the 15\% threshold for $p$-type transport at all three temperatures, in contrast to the $n$-type case, for which $V_{\mathrm{Se}}$ does not reach the threshold within the investigated dilute-defect range. The $p$-type $C_{\mathrm{crit}}$ values are substantially higher in $\beta$-SnSe than in $\alpha$-SnSe, indicating greater defect tolerance of hole transport in the high-temperature $\beta$ phase. Their temperature dependence is nonmonotonic: for $V_{\mathrm{Sn}}$, $C_{\mathrm{crit}}$ changes from $2.016\times10^{-3}$ at 800~K to $2.130\times10^{-3}$ at 900~K and then decreases to $1.738\times10^{-3}$ at 1000~K. Similarly, for $V_{\mathrm{Se}}$, it changes from $3.147\times10^{-3}$ to $3.270\times10^{-3}$ and then to $2.625\times10^{-3}$ over the same temperature range.

The corresponding $yy$-direction results are summarized in Table~S7 (Supporting Information). The overall trends are similar to those along $xx$, with $V_{\mathrm{Se}}$ generally more defect tolerant than $V_{\mathrm{Sn}}$ and $\beta$-SnSe more tolerant than $\alpha$-SnSe. The remaining directional differences tare consistent with the anisotropic balance between e-ph and e-d scattering, which changes the relative effect of vacancies on the phonon-limited conductivity and hence on $C_{\mathrm{crit}}$.

These results establish a defect-tolerance hierarchy across carrier type, phase, and vacancy species. $n$-type $\beta$-SnSe with $V_{\mathrm{Se}}$ remains the most defect-tolerant configuration considered here, because the 15\% degradation threshold is not reached within the investigated dilute-defect range. In contrast, $p$-type $\alpha$-SnSe with $V_{\mathrm{Sn}}$ at 100~K has the lowest defect-tolerance limit, with $C_{\mathrm{crit}}=8.841\times10^{-5}$. Together with the SSCHA-corrected power factors reported above, these $C_{\mathrm{crit}}$ values provide quantitative defect-concentration limits for maintaining the PF of monolayer SnSe.

The PF degradation quantified here provides the electronic contribution to future $ZT$ optimization. While vacancies reduce the PF through carrier scattering, they can also suppress the lattice thermal conductivity. In bulk $Pnma$-SnSe, both $V_{\mathrm{Sn}}$ and $V_{\mathrm{Se}}$ reduce $\kappa_{\mathrm{L}}$, with a stronger effect for $V_{\mathrm{Se}}$ \cite{zhang2024effects}. Combined with our finding that $V_{\mathrm{Se}}$ causes weaker electron-defect scattering than $V_{\mathrm{Sn}}$, this suggests that Se vacancies may offer a better trade-off between electrical and thermal transport. Direct calculations of vacancy-dependent $\kappa_{\mathrm{L}}$ in monolayer SnSe are still needed to assess their impact on $ZT$.

\section{Conclusions}
We combine SSCHA-renormalized phonons with mode-resolved electron-phonon and electron-defect transport calculations to determine how anharmonic lattice dynamics and intrinsic vacancies constrain the PF of monolayer $\alpha$- and $\beta$-SnSe. The harmonic spectrum of $\alpha$-SnSe is dynamically stable, so SSCHA acts as a finite-temperature phonon correction rather than as a stabilization mechanism. For the $n$-type response emphasized here, phonon renormalization leaves the qualitative PF scale and temperature trend unchanged while redistributing scattering among optical branches: TO-2 provides the largest resolved contribution at 100~K, whereas LO-1 dominates at 300~K.

The harmonic soft mode of $\beta$-SnSe is stabilized by finite-temperature anharmonic renormalization at 800-1000~K, allowing transport calculations for the \textit{Cmcm} phase. LO/TO-2 remains the dominant electron-scattering channel over this temperature range. At carrier densities near $10^{12}$~cm$^{-2}$, the $n$-type PF reaches 15-19~$\mu\mathrm{W}/(\mathrm{K}^{2}\cdot\mathrm{cm})$ at 800-900~K and is higher than the $p$-type PF mainly because of its larger electrical conductivity. The larger maxima near $10^{14}$~cm$^{-2}$ should be regarded as phonon-limited upper bounds, since such carrier densities are difficult to maintain at 800-1000~K and may also exceed the range where the rigid-band approximation remains reliable.

Vacancy scattering adds a strong carrier- and phase-dependent constraint. Sn vacancies cause shorter electron-defect relaxation times and greater PF degradation than Se vacancies, while $p$-type transport is generally more sensitive to defects than $n$-type transport. Using a 15\% PF-degradation criterion, the lowest critical concentration is $C_{\mathrm{crit}}=8.841\times10^{-5}$ for $p$-type $\alpha$-SnSe with $V_{\mathrm{Sn}}$ at 100~K. In contrast, for $n$-type $\beta$-SnSe with $V_{\mathrm{Se}}$, the 15\% threshold is not reached up to $5\times10^{-3}$. Overall, Sn vacancies impose the stricter limit on preserving the phonon-limited PF, particularly in $\beta$-SnSe.
\section{Computational Details}

\subsection{First-principles electronic structure} \label{sec:scf_nscf_computational_details}
We perform first-principles density functional theory (DFT) calculations for monolayer SnSe using the \textsc{Quantum ESPRESSO} package \cite{QE-2009, QE-2017}. All calculations use the Perdew-Burke-Ernzerhof (PBE) exchange-correlation functional, together with SSSP efficiency ultrasoft pseudopotentials (USPP) for both Sn and Se \cite{prandini2018precision, Garrity2014-ka}.

Monolayer SnSe is modeled within a supercell using a vacuum spacing of 20~\AA\ along the out-of-plane direction to suppress interlayer interactions. In addition, to eliminate residual long-range electrostatic coupling between periodic images of the slab, we employ the 2D Coulomb cutoff for all self-consistent (SCF) and non-self-consistent (NSCF) calculations \cite{PhysRevB.96.075448}. Both the primitive cell and all defect-containing supercells are relaxed using the Broyden-Fletcher-Goldfarb-Shanno (BFGS) algorithm, with convergence criteria of $\Delta E < 1.0\times10^{-6}$ Ry for the total energy and $|\mathbf{F}| < 1.0\times10^{-5}$ Ry/Bohr for the Hellmann-Feynman forces~\cite{hung2022quantum}. SCF convergence is achieved with a total-energy convergence criterion of $1.0\times10^{-9}$ Ry. A kinetic-energy cutoff of 70~Ry is used for the plane-wave expansion of the Kohn-Sham wave functions, together with a charge-density cutoff of 560~Ry.
Brillouin-zone sampling is performed using a $12\times12\times1$ Monkhorst-Pack $\mathbf{k}$-point grid for the primitive cell, whereas a $3\times3\times1$ grid is adopted for the $4\times4\times1$ supercell.

The NSCF calculation uses a uniform $24\times24\times1$ $\bm{k}$-point grid to obtain converged Bloch states for the primitive cell.
The resulting Bloch states are used to construct maximally localized Wannier functions (MLWFs) using Wannier90 \cite{Pizzi_2020, PhysRevB.65.035109}. We employ a 14-band Wannierization based on the projectors Sn: $s$, $p_x$, $p_y$, $p_z$ and Se: $p_x$, $p_y$, $p_z$. The MLWFs are used to interpolate band energies and velocities onto dense Brillouin-zone meshes for the transport calculations.

\subsection{Harmonic phonons and anharmonic renormalization (SSCHA)}
Harmonic phonon frequencies are computed within density-functional perturbation theory (DFPT), as implemented in \textsc{Quantum ESPRESSO} \cite{QE-2009, QE-2017}, using dynamical matrices calculated on a $5\times5\times1$ $\mathbf{q}$-point grid.
To include finite-temperature anharmonic effects, we employ the stochastic self-consistent harmonic approximation (SSCHA)~\cite{Monacelli2021-cx, Ribeiro_undated-sq, Ranalli2024-iu, Chang2025-xw} to obtain effective force constants at each temperature and, consequently, renormalized phonon frequencies. In SSCHA, an auxiliary harmonic Hamiltonian is optimized by minimizing the vibrational free energy \cite{Monacelli2021-cx}. The minimization is performed using total energies and Hellmann-Feynman forces computed for an ensemble of stochastic configurations sampled from the trial density matrix. These quantities are evaluated in $5\times5\times1$ supercells using the same SCF settings as in Sec.~\ref{sec:scf_nscf_computational_details}; Brillouin-zone sampling for the SSCHA supercells is performed using a $1\times1\times1$ $\mathbf{k}$-point grid. The stochastic ensemble comprises 100 configurations. The SSCHA cycle is stopped when the maximum change in the renormalized phonon frequencies falls below $2~\mathrm{cm^{-1}}$.

\subsection{Electron-phonon and electron-defect scattering}

The electrical transport coefficients are calculated within the Boltzmann transport equation (BTE) and the relaxation-time approximation (RTA) \cite{Mu-Mu-2025-cz, Hung2017-bj, Gupta2021-nm}. For each scattering branch, the transport distribution function (TDF) is
\begin{equation}
\Sigma(E,T)=\frac{1}{N_k}\sum_{n\bm{k}}
\mathbf{v}_{n\bm{k}}\otimes\mathbf{v}_{n\bm{k}}\,\tau_{n\bm{k}}(T)\,
\delta\!\left(E-\varepsilon_{n\bm{k}}\right),
\label{eq:TDF}
\end{equation}
where $\mathbf{v}*{n\bm{k}}=\hbar^{-1}\nabla*{\bm{k}}\varepsilon_{n\bm{k}}$ is the group velocity, $\varepsilon_{n\bm{k}}$ is the band energy, and $\tau_{n\bm{k}}$ is the carrier relaxation time. Electron-phonon and electron-defect scattering are treated separately using their respective relaxation times.

\subsubsection{Electron-phonon interaction}
The e-ph scattering rate is evaluated within lowest-order perturbation theory using Fermi's golden rule \cite{Zhou2021-hc, Bao2025-nl}. \begin{equation}
\begin{multlined}
\left(\tau_{n\bm{k}}^{\mathrm{e\text{-}ph}}\right)^{-1}
=
\frac{2\pi}{\hbar}\,\frac{1}{N_q}
\sum_{n'\nu\bm{q}}
\left|g^{\nu}_{n\bm{k},\,n'\bm{k}+\bm{q}}\right|^2
\\
\times \Bigl[
\left(n_{\nu\bm{q}}+f^{0}_{n',\bm{k}+\bm{q}}\right)
\delta\!\left(\Delta\varepsilon_{nn'}(\bm{k},\bm{q})+\hbar\omega_{\nu\bm{q}}\right)
\\
+\left(1+n_{\nu\bm{q}}-f^{0}_{n',\bm{k}+\bm{q}}\right)
\delta\!\left(\Delta\varepsilon_{nn'}(\bm{k},\bm{q})-\hbar\omega_{\nu\bm{q}}\right)
\Bigr].
\end{multlined}
\label{eq:eph_rate}
\end{equation}
In Eq.~\eqref{eq:eph_rate}, $N_q$ is the number of $\bm{q}$ points used in the summation, and
$\bm{q}$ and $\nu$ are the phonon wavevector and branch, with the final electronic state
$\bm{k}'=\bm{k}+\bm{q}$.
Here and below, $\varepsilon_{n\bm{k}}$ and $\hbar\omega_{\nu\bm{q}}$ are the electron and phonon
energies, $f^{0}_{n',\bm{k}+\bm{q}}$ and $n_{\nu\bm{q}}$ are the equilibrium Fermi-Dirac and
Bose-Einstein occupations, and
$\Delta\varepsilon_{nn'}(\bm{k},\bm{q})=\varepsilon_{n',\bm{k}+\bm{q}}-\varepsilon_{n\bm{k}}$;
the two $\delta$-terms correspond to phonon emission and absorption, respectively.

The electron-phonon coupling (EPC) matrix element in Eq.\eqref{eq:eph_rate} is defined as
\begin{equation}
g^{\nu}_{n\bm{k},\,n'\bm{k}+\bm{q}}
=
\left\langle
\psi_{n',\bm{k}+\bm{q}}
\middle|
V_{\nu\bm{q}}
\middle|
\psi_{n,\bm{k}}
\right\rangle,
\label{eq:epc_me}
\end{equation}
where $\psi_{n,\bm{k}}$ is the Kohn-Sham Bloch state and $V_{\nu\bm{q}}$ is the first-order
variation of the self-consistent potential associated with the phonon perturbation of mode
$(\nu,\bm{q})$, obtained within density-functional perturbation theory (DFPT). Electron-phonon scattering rates are computed using the \textsc{Perturbo} code (version 3.0.1) with the 2D polar electron-phonon correction enabled \cite{Zhou2021-hc}.
Electron eigenvalues, band velocities, DFPT dynamical matrices, and the corresponding first-order perturbation potentials are then interpolated onto dense $200\times200\times1$ $\bm{k}$ and $100\times100\times1$ $\bm{q}$ meshes to obtain converged e-ph scattering rates.

\subsubsection{Electron-defect interaction}

For electron-defect (e-d) scattering, the state-resolved relaxation rate is evaluated as
\begin{equation}
\left(\tau^{\mathrm{e\text{-}d}_{n\bm{k}}}\right)^{-1}
=
\frac{2\pi}{\hbar}
\frac{n_{\mathrm{at}} C_d}{N_{\bm{k}'}}
\sum_{n'\bm{k}'}
\left|M_{n'\bm{k}',\,n\bm{k}}\right|^2
\delta\!\left(
\varepsilon_{n'\bm{k}'}-\varepsilon_{n\bm{k}}
\right),
\label{eq:ed_rate}
\end{equation}
where $C_d$ is the defect concentration, $n_{\mathrm{at}}$ is the
number of atoms in the primitive cell, and $N_{\bm{k}'}$ is the
number of $\bm{k}'$ points used to sample the final electronic states
in the Brillouin zone. Here, $\varepsilon_{n\bm{k}}$ is the electronic energy and $\delta(\cdot)$ enforces energy conservation for elastic scattering.

The e-d matrix element is defined as
\begin{equation}
M_{n'\bm{k}',\,n\bm{k}}
=
\left\langle \psi_{n'\bm{k}'} \middle| \Delta V_{d} \middle| \psi_{n\bm{k}} \right\rangle,
\qquad
\Delta V_{d}=V_{\mathrm{KS}}^{(d)}-V_{\mathrm{KS}}^{(p)},
\label{eq:M_def}
\end{equation}
where $\psi_{n\bm{k}}$ is the Kohn-Sham Bloch state, and $V_{\mathrm{KS}}^{(d)}$ and
$V_{\mathrm{KS}}^{(p)}$ are the Kohn-Sham potentials of the defect-containing and pristine
systems, respectively. The explicit derivation of Eqs.~\eqref{eq:ed_rate}-\eqref{eq:M_def} and
the associated normalization conventions
follow the defect-scattering formalism of \citeauthor{Lu2019-ch}~\cite{Lu2019-ch}.

In the dilute-defect limit, we approximate the electronic band structure entering Eq.~\eqref{eq:ed_rate} by that of the pristine crystal, while the scattering is driven by the defect-induced perturbation potential $\Delta V_d$. The e-d scattering rates are computed using the electron-defect branch of \textsc{Perturbo} \cite{Lu2019-ch, Lu2020-cv}. The perturbation potential $\Delta V_d$ is constructed from a pair of $4\times4\times1$ supercell calculations (defect-containing and pristine), with the defect placed at the supercell center. Brillouin-zone summations are performed on a dense $200\times200\times1$ mesh.

\subsection{Combination of Scattering Contributions to Thermoelectric Transport}

Thermoelectric coefficients are evaluated within the BTE formalism using the transport distribution function $\Sigma(E,T)$ defined in Eq.~\eqref{eq:TDF}. The generalized transport integrals are
\begin{equation}
\mathcal{L}^{(\alpha)}(T,\mu)
=
e^{2}\int dE\,\Sigma(E,T)\,(E-\mu)^{\alpha}
\left(-\frac{\partial f_{0}(E,\mu,T)}{\partial E}\right),
\label{eq:Lalpha}
\end{equation}
from which the electrical conductivity $\sigma$ and Seebeck coefficient $S$ are obtained as
\begin{equation}
\sigma=\mathcal{L}^{(0)},
\label{eq:sigma_def}
\end{equation}
\begin{equation}
S=-\frac{1}{eT}\frac{\mathcal{L}^{(1)}}{\mathcal{L}^{(0)}},
\label{eq:S_def}
\end{equation}
where $e>0$ is the elementary charge.

For e-ph scattering, $\sigma^{\mathrm{e\text{-}ph}}$ and $S^{\mathrm{e\text{-}ph}}$ are obtained directly from \textsc{Perturbo} (version 3.0.1) \cite{Zhou2021-hc}. For e-d scattering, the present e-d workflow provides state-resolved relaxation times $\tau^{\mathrm{e\text{-}d}}_{n\bm{k}}$ (Eq.~\eqref{eq:ed_rate}) but does not output the Seebeck coefficient. Therefore, $S^{\mathrm{e\text{-}d}}$ is computed by post-processing the same electronic structure with \textsc{BoltzTraP2}, using $\tau^{\mathrm{e\text{-}d}}_{n\bm{k}}(T)$ from the \textsc{Perturbo} as an external relaxation-time input \cite{Madsen2018-rw}.

We combine e-ph and e-d scattering at the level of the transport observables rather than by matching state-resolved relaxation times. The two branches are generated using different \textsc{Quantum ESPRESSO} versions, which produce small numerical differences in $\varepsilon_{n\bm{k}}$ and $\bm{v}_{n\bm{k}}$ on the ultradense meshes. Direct one-to-one matching of $(n,\bm{k})$ states is therefore unreliable. We instead treat e-ph and e-d interactions as independent scattering mechanisms evaluated at the same chemical potential $\mu$ and temperature $T$.

The electrical conductivity is combined by adding the partial resistivities,

\begin{equation}
\rho \equiv \sigma^{-1}
=
\left(\sigma^{\mathrm{e\text{-}ph}}\right)^{-1}
+
\left(\sigma^{\mathrm{e\text{-}d}}\right)^{-1}.
\label{eq:rho_add_final}
\end{equation}

Under the same assumption of independent scattering, the Seebeck coefficient is combined using the Nordheim-Gorter rule \cite{mortensen1979transport, Hinterleitner2019-ck, PhysRev.138.A105},

\begin{equation}
S =
\frac{
\rho^{\mathrm{e\text{-}ph}}S^{\mathrm{e\text{-}ph}}
+
\rho^{\mathrm{e\text{-}d}}S^{\mathrm{e\text{-}d}}
}
{
\rho^{\mathrm{e\text{-}ph}}
+
\rho^{\mathrm{e\text{-}d}}
}
=
\frac{
\dfrac{S^{\mathrm{e\text{-}ph}}}{\sigma^{\mathrm{e\text{-}ph}}}
+
\dfrac{S^{\mathrm{e\text{-}d}}}{\sigma^{\mathrm{e\text{-}d}}}
}
{
\dfrac{1}{\sigma^{\mathrm{e\text{-}ph}}}
+
\dfrac{1}{\sigma^{\mathrm{e\text{-}d}}}
}.
\label{eq:S_rho_weighted_final}
\end{equation}
\section*{CRediT authorship contribution statement}
\textbf{Nguyen Tran Gia Bao:} Writing - original draft, Visualization, Data curation, Methodology, Investigation, Formal analysis. \textbf{Thang Bach Phan:} Resources, Funding acquisition. \textbf{Vu Thi Hanh Thu:} Supervision, Writing - review and editing, Funding acquisition, Visualization, Investigation. \textbf{Nguyen Tuan Hung:} Supervision, Writing - review and editing, Validation, Methodology, Investigation, Formal analysis.

\section*{Declaration of competing interest}
The authors declare that they have no known competing financial interests or personal relationships that could have appeared to influence the work reported in this paper.

\section*{Acknowledgements}
This research is funded by Ho Chi Minh City Department of Science and Technology, Vietnam under contract number 25/2025/HD-QKHCN; through the Project for Developing Mechanisms to Foster the Establishment and Growth of International-Standard Centers of Excellence (CoE Project) as per Decision No. 5721/QD-UBND dated December 11, 2023, of the People Committee of Ho Chi Minh City.


\section*{Supplementary Information}
Supporting Information associated with this article includes additional structural, electronic, phonon, and transport data.
\appendix
\section{\label{sec:appendix_Cd}Defect-concentration scaling of transport coefficients}

The electron-defect scattering rate scales linearly with the defect concentration $C_d$:
\begin{equation}
\left(\tau^{\mathrm{e\text{-}d}}_{n\bm{k}}\right)^{-1}
=
C_d\,\Gamma_{n\bm{k}},
\label{eq:tau_ed_linear}
\end{equation}
where $\Gamma_{n\bm{k}}$ is determined by the pristine electronic structure and the defect-induced perturbation potential $\Delta V_d$ and is independent of $C_d$. Here, $C_d$ is the dimensionless defect fraction, and $C_d^{\mathrm{ref}}=10^{-6}$ (1~ppm) is the reference concentration used in the electron-defect calculations. Consequently, the relaxation time at any defect concentration $C_d$ is obtained as
\begin{equation}
\tau^{\mathrm{e\text{-}d}}_{n\bm{k}}(C_d)
=
\frac{C_d^{\mathrm{ref}}}{C_d}
\tau^{\mathrm{e\text{-}d}}_{n\bm{k}}
\left(C_d^{\mathrm{ref}}\right).
\label{eq:tau_scaling}
\end{equation}

This prefactor propagates through the transport distribution function (Eq.~\eqref{eq:TDF}) and the transport integrals (Eq.~\eqref{eq:Lalpha}). The defect-limited conductivity and Seebeck coefficient therefore obey
\begin{align}
\sigma^{\mathrm{e\text{-}d}}(C_d)
&=
\frac{C_d^{\mathrm{ref}}}{C_d}\,\sigma_{\mathrm{ref}}^{\mathrm{e\text{-}d}},
\label{eq:sigma_ed_scaling}\\
S^{\mathrm{e\text{-}d}}(C_d)
&=S_{\mathrm{ref}}^{\mathrm{e\text{-}d}},
\label{eq:S_ed_const}
\end{align}
where $\sigma_{\mathrm{ref}}^{\mathrm{e\text{-}d}}$ and $S_{\mathrm{ref}}^{\mathrm{e\text{-}d}}$ are evaluated at 1~ppm. The uniform concentration prefactor cancels from the ratio $\mathcal{L}^{(1)}/\mathcal{L}^{(0)}$, so the defect-limited Seebeck coefficient is independent of $C_d$ within this dilute-defect scaling.

Combining $\sigma^{\mathrm{e\text{-}ph}}$ and $\sigma^{\mathrm{e\text{-}d}}(C_d)$ through Eq.~\eqref{eq:rho_add_final} gives
\begin{equation}
\sigma^{\mathrm{e\text{-}ph+e\text{-}d}}(C_d)
=
\frac{
\sigma^{\mathrm{e\text{-}ph}}\,\sigma_{\mathrm{ref}}^{\mathrm{e\text{-}d}}
}{
\sigma_{\mathrm{ref}}^{\mathrm{e\text{-}d}}
+
\dfrac{C_d}{C_d^{\mathrm{ref}}}\sigma^{\mathrm{e\text{-}ph}}
}.
\label{eq:sigma_total_Cd}
\end{equation}
The Nordheim-Gorter rule (Eq.~\eqref{eq:S_rho_weighted_final}) yields
\begin{equation}
S(C_d)
=
\frac{S^{\mathrm{e\text{-}ph}}+\alpha S_{\mathrm{ref}}^{\mathrm{e\text{-}d}}}{1+\alpha},
\qquad
\alpha
\equiv
\frac{C_d}{C_d^{\mathrm{ref}}}
\frac{\sigma^{\mathrm{e\text{-}ph}}}{\sigma_{\mathrm{ref}}^{\mathrm{e\text{-}d}}}.
\label{eq:S_total_Cd}
\end{equation}
The calculated $S^{\mathrm{e\text{-}ph}}$ and $S_{\mathrm{ref}}^{\mathrm{e\text{-}d}}$ values are similar over the relevant carrier-concentration windows (Fig.~S19, Supporting Information). With the approximation $S^{\mathrm{e\text{-}ph}}\simeq S_{\mathrm{ref}}^{\mathrm{e\text{-}d}}\equiv S$, Eq.~\eqref{eq:S_total_Cd} reduces to $S(C_d)\simeq S$, and
\begin{equation}
\mathrm{PF}^{\mathrm{e\text{-}ph+e\text{-}d}}(C_d)
=
\frac{
S^{2}\sigma^{\mathrm{e\text{-}ph}}\,\sigma_{\mathrm{ref}}^{\mathrm{e\text{-}d}}
}{
\sigma_{\mathrm{ref}}^{\mathrm{e\text{-}d}}
+
\dfrac{C_d}{C_d^{\mathrm{ref}}}\sigma^{\mathrm{e\text{-}ph}}
}.
\label{eq:PF_Cd}
\end{equation}

\section*{Data availability}
Data available on request.
\bibliographystyle{elsarticle-num-names} 
\bibliography{references}

\end{document}